\documentclass[american,aps,pra,reprint, superscriptaddress]{revtex4-1}
\usepackage{color}
\usepackage{times}
\usepackage{subfig}
\usepackage{bm}
\usepackage{graphicx}
\usepackage{graphics}
\DeclareGraphicsExtensions{.pdf,.png,.jpg}
\usepackage{amsbsy}
\usepackage{amsmath}
\usepackage{amsfonts}
\usepackage{amsthm}
\usepackage{float}
\usepackage{amssymb}
\usepackage{textcomp}
\usepackage{mathtools}
\usepackage[subnum]{cases}
\usepackage{multirow}

\DeclareMathOperator{\Tr}{Tr}
\definecolor{purple(html/css)}{rgb}{0.5, 0.0, 0.5}
\newcommand{\ket}[1]{| #1 \rangle}
\newcommand{\bra}[1]{\langle #1 |}

\newcommand{\ketbra}[2]{| #1 \rangle \langle #2 |}

\newcommand{\me}{\mathrm{e}}
\newcommand{\mi}{\mathrm{i}}
\begin{document}

\title{Coherence of quantum channels}
\author{Chandan Datta}
\email{chandan@iopb.res.in}
\affiliation{Institute of Physics, Sachivalaya Marg,
Bhubaneswar 751005, Odisha, India.}
\affiliation{Homi Bhabha National Institute, Training School Complex, Anushakti Nagar, Mumbai 400085, India.}

\author{Sk Sazim}
\email{sk.sazimsq49@gmail.com}
\affiliation{QIC group, Harish-Chandra Research Institute, HBNI, Allahabad 211019, India}

\author{Arun K. Pati}
\email{akpati@hri.res.in}
\affiliation{QIC group, Harish-Chandra Research Institute, HBNI, Allahabad 211019, India}

\author{Pankaj Agrawal}
\email{agrawal@iopb.res.in}
\affiliation{Institute of Physics, Sachivalaya Marg,
Bhubaneswar 751005, Odisha, India.}
\affiliation{Homi Bhabha National Institute, Training School Complex, Anushakti Nagar, Mumbai 400085, India.}

\begin{abstract}
We investigate the coherence of quantum channels using the
Choi-Jamio\l{}kowski isomorphism.  The relation between the coherence and
the purity of the channel respects a duality relation. It characterizes
the allowed values of coherence when the channel has certain purity. This
duality has been depicted via the Coherence-Purity (Co-Pu) diagrams. In
particular, we study the quantum coherence of the unital and non-unital
qubit channels and find out the allowed region of coherence for a fixed
purity. We also study coherence of different incoherent channels, namely,
incoherent operation (IO), strictly incoherent operation (SIO), physical
incoherent operation (PIO) etc. Interestingly, we find that the allowed
region for different incoherent operations maintain the relation
$PIO\subset SIO \subset IO$. In fact, we find that if PIOs are coherence
preserving operations (CPO), its coherence is zero otherwise it has unit
coherence and unit purity. Interestingly, different kinds of qubit
channels can be distinguished  using the Co-Pu diagram. 
The unital channels generally do not create
coherence whereas some nonunital can.  All
coherence breaking channels are shown to have zero coherence, whereas,
this is not usually true for entanglement breaking channels.  It turns out
that the coherence preserving qubit channels have unit coherence. 
Although the coherence of the Choi matrix of the incoherent channels might
have finite values, its subsystem contains no coherence. This indicates
that the incoherent channels can either be unital or nonunital under some
conditions.  
\end{abstract}

\maketitle
\section{Introduction}
Quantum coherence and entanglement are two fundamental resources in quantum information and computation  \cite{entang, entangr, cohereR, cohereR1}. 
These two resources are closely related \cite{cohere1,cohere2E}. While the concept of entanglement requires at least two particles, 
coherence can be defined for a single system. Recent developments show that coherence
in a quantum system can be a useful resource in quantum algorithm 
\cite{fan-app, pati-app,raste-app, hilari-app, matera-app}, quantum meteorology \cite{coh-metro}, 
quantum thermodynamics \cite{cohere6,cohere7,cohere8,cohere9,cohereC1,cohereC2,cohereC3}, 
and quantum biology \cite{cohere3,cohere4,cohere5}. Therefore, the study of resource theory of quantum coherence is of immense importance 
\cite{cohereC4,cohereC5,cohereC6,cohereC7,cohereC8,cohereC9,cohereD1,cohereD2,cohereD3,cohereD4,cohereD5, 
cohereD6,cohereD7,cohereD8,cohereD9,cohereA1,cohereA2,cohereA3,cohereA4,cohereA5,cohereA6,cohereA7,cohereA8,cohereA9, 
cohereB1,cohereB2,cohereB4,cohereB5,cohereB6,chiru, non-orth}.

The quantum coherence like other quantum resources is also fragile in the presence of noisy
 environment. The interaction of quantum systems with environment have  been extensively 
 studied using different models -- in particular using noisy channels \cite{cohereB7}. 
Characterizing all these channels and their effect on various physical resources are vital  
\cite{cohereB7,cohereB8}. These channels are also important to construct resource theoretic aspect of coherence. Here, in this work, we ask a 
reverse question. Can we associate coherence with 
a quantum channel? We answer this question positively. We define the coherence of quantum 
channels using the Choi-Jamio\l{}kowski (C-J) isomorphism \cite{cohereB9,cohereE1}. 
In this paper, we consider the unital as well as non-unital qubit channels \cite{prescC1,prescC2,cohereB8}. 
We compute their coherence and purity analytically. While coherence of a non-unital channel can 
go up to $\sqrt{2}$ as measured by the $l_1$-norm, the coherence of unital channels can never exceed $1$. Using the coherence-purity (CoPu) 
diagrams, we find that it may be possible to distinguish unital channels and non-unital channels. 

The resource theory of coherence require two important elements -- free states and free operations \cite{cohereC4,cohereR}. Free states are those which have
no coherence in a given reference basis. Free operations do not create any coherence and are known as incoherent operations. Depending on the restrictions 
(physical requirement), there exist different types of incoherent operations. The largest set of incoherent operations contains
Maximally Incoherent Operations (MIO) \cite{mio-aberg}. The other candidates are Incoherent Operations (IO) \cite{cohereC4}, Strictly Incoherent Operations 
(SIO) \cite{cohereR1,cohereC5}, Physical 
Incoherent Operations (PIO) \cite{cohereC9} etc. There are many other free operations in the literature like Fully Incoherent Operations (FIO), 
Genuine Incoherent Operations (GIO) etc. It is an important task to understand these operations and distinguish them. In this work, we aim to distinguish these operations 
using CoPu diagrams. 
A PIO is in fact a strange candidate -- 
some PIOs which are not Coherence Preserving Operations (CPO) \cite{cohereA1} have coherence zero but otherwise it has unit purity and unit coherence. We have constructed 
all possible FIOs for qubit case and show that they have zero coherence when only row elements are nonzero in their Kraus representation. The FIOs which are 
diagonal are called GIOs \cite{pio-eric2}. The FIOs with anti-diagonal elements in their Kraus representation have same CoPu diagrams as GIOs.

We also consider the class of coherence non-generating qubit channels (CNC) as well as the channels to create maximal coherence (CMC). CNC is the bigger set 
in comparison to all incoherent 
operations \cite{Cngc}. We also consider other known qubit channels like the class of Pauli channels, degradable and anti-degradable channels, 
amplitude damping channels, depolarizing channels, and homogenization channels and show that they might be distinguished using CoPu diagrams.
Following are the salient features of our results which we address extensively 
in the main text.

\begin{itemize}

\item The coherence of quantum channels has been identified with the coherence of the Choi matrix. 
The characterization of coherence quantifies the quantumness of the channels, i.e., it might be 
considered as the quantity which characterizes how much the map is quantum.

\item The relation between the coherence and the purity of the channel respects a duality type 
relation. It characterizes the allowed values of coherence while the channel has certain purity. 
This duality has been depicted via the Coherence-Purity (Co-Pu) diagrams.

\item Different kinds of qubit channels can be distinguished 
using the Co-Pu diagrams. For example, the unital and nonunital channels, the incoherent channels, degradable 
and anti-degradable channels etc can be distinguished with the help of our formalism.

\item  Unital channels generally do not create coherence
 whereas some nonunital can. 

\item  a) All coherence breaking channels have zero coherence. However, this is not usually true for 
entanglement breaking channels. b) Moreover, the coherence preserving qubit channels have unit coherence.

\item Although the coherence of the Choi matrix of the incoherent channels might have finite values, its subsystem contains no coherence. This very fact tells us that the incoherent channels can either be unital or nonunital with $\vec{\tau} =\{0,0,\tau_z\}$. ($\vec{\tau}$ is defined in Section II-A.)

\end{itemize}

The paper is organized as follows. In Section II, we introduce some extant concepts relevant to our work. 
Section III contains the detailed analysis of 
coherence content of all qubit unital as well as non-unital channels. Section IV presents the CoPu diagrams of the incoherent operations introduced recently 
in the literature. In Section V, we discuss CoPu diagrams of other relevant qubit channels. Finally, we conclude in the last section.

\section{Preliminaries}
Here, we will discuss some relevant concepts which are important in explaining our main results.
\subsection{Quantum Channels}
A quantum channel is a completely positive and trace preserving (CPTP) linear map which  
maps a density 
matrix to a density matrix \cite{cohereB7}. If a map, $\Phi$, is CPTP, then it can be represented by a set of 
Kraus operators $\{K_i;i=1,2,...,n\}$
\begin{equation}\label{kraus rep}
\Phi [\rho] : =\sum_i^n K_i\rho K_i^{\dagger},
\end{equation}
with $\sum_i^nK_i^{\dagger}K_i=1$. Here $\rho$ is an arbitrary density matrix.
We call this representation as the Kraus representation of the channel (KROC). 
However, in this work, we will mainly focus on qubit channels \cite{prescC1,prescC2}. 

The action of a qubit channel  
$\Phi$ can also be completely characterized by a $3 \times 3$ real matrix $M$ and a $3$-dimensional vector 
$\vec{\tau}$ \cite{prescC1,prescC2,cohereB8}. An arbitrary qubit is expressed as  
$\rho=\frac{1}{2}(\mathbb{I}+\vec{r}\cdot\vec{\sigma})$, where $\vec{r}$ is the $3$-dimensional Bloch vector. 
The action of qubit channel $\Phi$ on $\rho$ is described in following way: 
\begin{equation}
(1, \vec{r}^\prime)^T=\Lambda_{\Phi}(1, \vec{r})^T,
\end{equation}
where $\Lambda_{\Phi}$ represents a real $4 \times 4$ matrix and $T$ denotes transposition. 
The most general form of $\Lambda_{\Phi}$ for 
complete positivity can be written as 
\begin{equation}
\Lambda_{\Phi}=\begin{pmatrix}
1 & 0_{1\times3} \\
\vec{\tau} & M
\end{pmatrix}.
\end{equation}
It leads to the affine transformation of
the Bloch vector, i.e., $\vec{r}^\prime=M\vec{r}+\vec{\tau}$. Up to some local unitary 
equivalence, any qubit channel can be written as 
\begin{equation}
\Lambda_{\Phi}=\begin{pmatrix}
1 & 0 & 0 & 0 \\
\tau_x & \lambda_x & 0 & 0 \\
\tau_y & 0 & \lambda_y & 0 \\
\tau_z & 0 & 0 & \lambda_z \\
\end{pmatrix},
\end{equation}
where $\lambda$'s are the (signed) singular values of the matrix $M$ and $\tau$'s represents the shift of 
the coordinates \cite{prescC1,prescC2}. A channel is unital if and only if $\vec{\tau}=0$. 

The above representation of a qubit channel $\Phi$ is known as the affine representation of the channel (AROC).

\subsection{Choi-Jamio\l{}kowski Isomorphism}
In this paper, we wish to characterize  a quantum channel by its coherence. 
To do this, we will use the idea of Choi-Jamio\l{}kowski isomorphism \cite{cohereB9,cohereE1}. 
It permits one to associate a CPTP map $\Phi$ to a density matrix of composite system $AB$ with $B$ being 
the auxiliary system of same dimension as $A$. The prescription is:
\begin{equation}\label{choi_jamilowski}
\rho_{AB}=\Phi \otimes\mathbb{I}_B (\ket{\Psi}_{AB}\bra{\Psi}), 
\end{equation}
where $\ket{\Psi}_{AB}$ is a maximally entangled state. It states 
that for every quantum state there is a unique quantum operation. It is also known as the channel-state 
duality. As there is one to one map between the state and the channel, the coherence of the final 
state $\rho_{AB}$ can represent the coherence of the quantum channel. If $\rho_{AB}$ is 
separable then the channel, $\Phi$ is called entanglement breaking channel. If the state $\rho_{AB}$ is 
incoherent, the corresponding channel is called coherence breaking channel. The density matrix $\rho_{AB}$ is called the Choi matrix.

For qubit channels, without loss of generality, we will consider the singlet state as the two qubit maximally entangled state. The canonical form of singlet state is
\begin{equation}\label{singlet state}
\ket{\Psi}_{AB}\bra{\Psi}=\frac{1}{4}(\mathbb{I}\otimes \mathbb{I}-\sum_{i=1}^3\sigma_i\otimes\sigma_i),
\end{equation}
where $\sigma_i$ ($i=1,2,3$) are the Pauli matrices. Coherence of a state depends on the 
reference basis used to write it. Here and below, we shall use computational basis as 
reference basis.
\subsection{Quantum coherence}
Quantum coherence is a fundamental concept in quantum mechanics. It arises due to the superposition principle \cite{mio-aberg}. 
Recently, much attention has been paid to define proper measure of quantum coherence \cite{cohereC4, cohereC9, cohereD7, 
pio-eric2}. 
As coherence is a basis dependent quantity, 
we should first fix a particular basis. Let $\{\ket{i}\}$ ($i=1\ldots d$) is a basis in a $d$-dimensional 
Hilbert space $\mathcal{H}_d$. The density matrices which are diagonal in this basis 
are called incoherent states. The structure of these density matrices is as follows
\begin{equation}\label{incoherent_state}
\delta=\sum_{i=1}^d \delta_i \ketbra{i}{i},
\end{equation}
where $\sum_{i=1}^d \delta_i=1$. 
Quantum operations with 
Kraus operators, $\{K_i\}$, satisfying $\sum_{i} K_i^\dagger K_i=\mathbb{I}$, 
will be incoherent if it takes an incoherent state to another incoherent state, i.e., 
$K_i\mathcal{I}K_i^\dagger \in \mathcal{I}$ for all $i$, where $\mathcal{I}$ is the set of all incoherent states.

Any proper measure of the coherence $C$ must satisfy the following conditions \cite{cohereC4,cohereR}:\\
\textbf{(C1)} $C(\delta)=0$, where $\delta \in \mathcal{I}$.  
Hence, for any quantum state $C(\rho)\geqslant 0$.\\
\textbf{(C2)} It should not increase under any incoherent operation, i.e., 
$C(\rho)\geqslant C(\Phi[\rho])$, where $\Phi[\rho]$ is any incoherent operation.\\
\textbf{(C3)} $C(\rho)$ is nonincreasing under selective measurements on average, 
$C(\rho)\geqslant\sum_iq_iC(\rho_i)$, where $q_i=\mbox{Tr}(K_i \rho K^\dagger_i)$ 
and $\rho_i=K_i \rho K^\dagger_i/q_i$ for all $i$ with $\sum_{i} K_i^\dagger K_i=\mathbb{I}$ 
and $K_i\mathcal{I}K_i^\dagger \in \mathcal{I}$.\\
\textbf{(C4)} $C(\rho)$ does not increase under mixing of quantum states, $\sum_ip_iC(\rho_i)\geqslant C(\sum_ip_i\rho_i)$, 
with $\rho=\sum_ip_i\rho_i$.

The $l_1$-norm of coherence, $C_{l_1}(\rho)$, 
and the relative entropy of coherence, $C_r(\rho)$, satisfy all these conditions. The $l_1$-norm of 
coherence measure is defined as 
\begin{equation}\label{l_1_norm}
C_{l_1}(\rho)=\sum_{i \neq j}|\rho_{i,j}|,
\end{equation} 
where $\rho_{i,j}=\bra{i}\rho\ket{j}$
and the relative entropy of coherence is defined as
\begin{equation}\label{relative entropy}
C_r(\rho)=S(\rho^D)-S(\rho),
\end{equation}
where $\rho^D$ is the matrix constructed from $\rho$ by removing all the off-diagonal elements and $S(\rho)$ represents the von Neumann entropy 
for the density operator $\rho$. 
This measure of quantum coherence satisfies all the criteria as required to be a good 
measure of coherence \cite{cohereR}. 

\section{Coherence of the channels}
Here we investigate the coherence of the unital as well as non-unital channels. According to C-J isomorphism, the coherence of the channels 
is equivalent to the coherence of the transformed singlet state. Hence, the channel coherence and its other properties can 
easily be evaluated. In this section, we will mainly follow the AROC. 
\subsection{Coherence of Unital Channels}
\begin{figure}[h]
\centering
\includegraphics[scale=0.5]{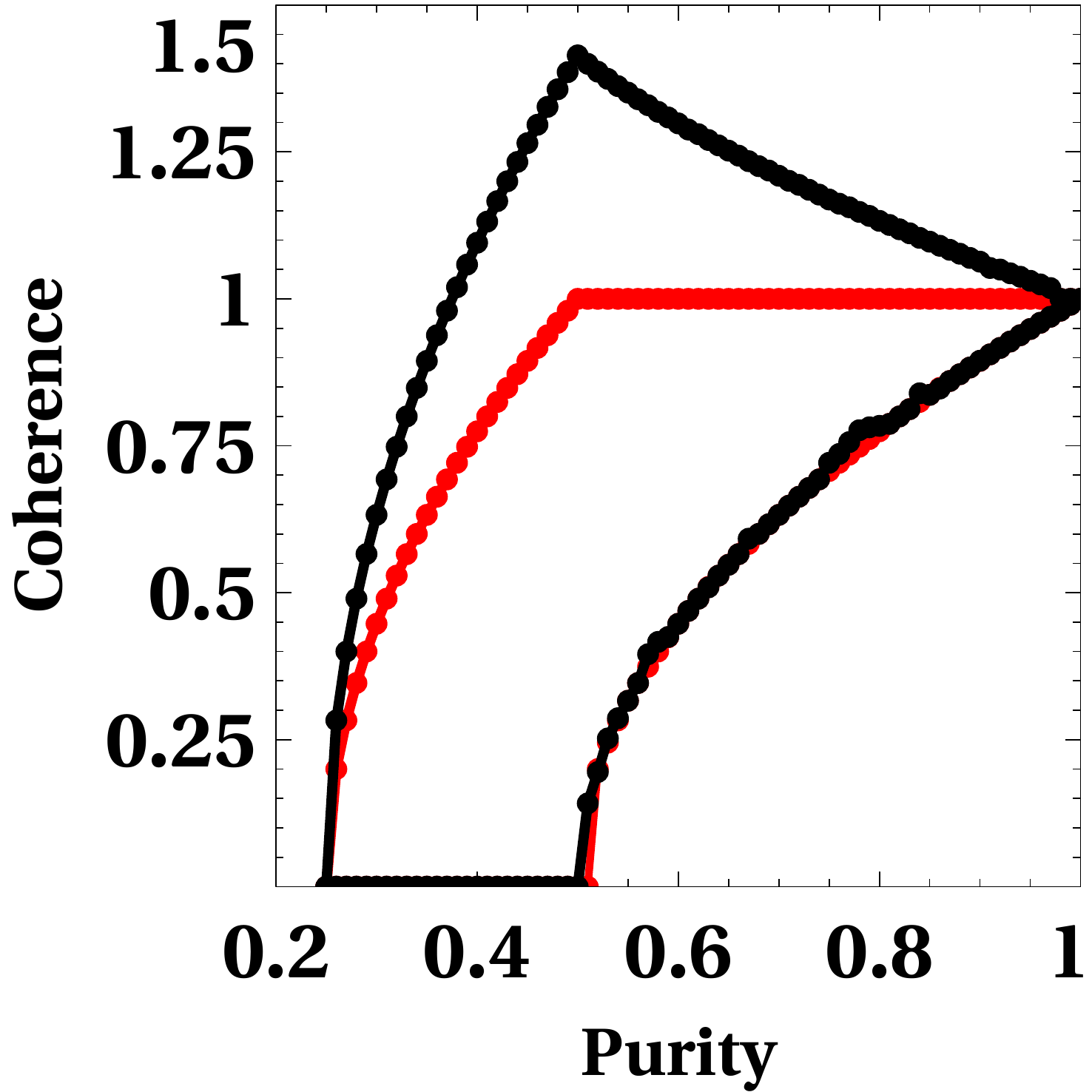}
\caption{(Color online) The figure depicts the allowed region of coherence as measured by the $l_1$-norm for unital 
(region inside red curve) and non-unital (region inside black curve) channels, respectively for the allowed purity range. The CoPu 
diagram shows that the channels outside the overlap region are non-unital and can be exactly distinguished from unital ones. The purity for  
these channels, $\mathcal{P}\in[\frac{1}{4},1]$.}
\label{unital-nonunital-reg}
\end{figure}

Unital qubit channels are those which do not change the maximally mixed state, $\mathbb{I}/2$. 
They satisfy, $\sum_i K_i^{\dagger}K_i=\mathbb{I}=\sum_i K_iK_i^{\dagger}$.  From the 
section II.B, it is clear that the set of unital channels can be represented as a three-parametric 
family of completely positive maps. Now if we apply the C-J map on the state given in 
(\ref{singlet state}), then the final state will be 
\begin{equation}\label{unital state}
\rho_{AB}= \frac{1}{4}\Big(\mathbb{I}\otimes \mathbb{I}-\vec{\lambda}\cdot(\vec{\sigma}\otimes \vec{\sigma})\Big).
\end{equation}
The positivity of the eigenvalues of $\rho_{AB}$ will ensure the complete positivity of the unital map. Let us define 
$q_{ij}=1+(-1)^{i}\lambda_x+(-1)^{i+j}\lambda_y+(-1)^{j}\lambda_z$ with $i,j=0,1$, where $q_{ij}$ are the four eigenvalues of the density matrix $\rho_{AB}$.
Therefore, the positivity constraints on the unital channels are \cite{cohereB8}
\begin{equation}\label{unitalcondi}
 q_{ij}\geq 0.
\end{equation}
Using the $l_1$-norm, the coherence of the unital channel is given by
\begin{equation}\label{unital coherence}
C_{l_1}=\frac{1}{2}\big(|\lambda_x+\lambda_y|+|\lambda_x-\lambda_y|\big).
\end{equation}
Note that the coherence does not depend on $\lambda_z$. It implies that many isocoherence 
planes will lie along $\lambda_z$ axis. This is the consequence of the choice
of reference basis.
The coherence of unital channel will reach its maximum value $1$ when 
we have $\lambda_x=\lambda_y=\pm 1$ or $\lambda_x=-\lambda_y=\pm 1$.
The purity, as defined by $\mathcal{P}=\rm{Tr}[\rho^2]$, of the unital channel is given by
\begin{equation}\label{unital purity}
\mathcal{P}=\frac{1}{4}(1+|\vec{\lambda}|^2).
\end{equation}
It is well known that the unital channels form a tetrahedron with unitary operators in its vertices \cite{cohereE2}. 
The state $\rho_{AB}$ has the same geometrical picture. Bell states sits on the four extremal 
points of the tetrahedron \cite{cohereE2}. From eq. (\ref{unital purity}) 
it is clear that $|\vec{\lambda}|^2=4\mathcal{P}-1$. Values of $\lambda_i$ lie 
on the surface of sphere with the radius 
$4\mathcal{P}-1$ centered at the point $\lambda_x=\lambda_y=\lambda_z=0$. The channels with the 
same purity form a sphere. Therefore, there may exist quantum channels with 
different coherence but the same purity (see Fig.\ref{unital-nonunital-reg}). 
\subsection{Coherence of Non-unital Channels}
The non-unital qubit channels are characterized by six parameters as shown in the section II-A. The Choi matrix 
corresponding to the non-unital channels is given by
\begin{equation}\label{nonunital state}
\rho_{AB}=\frac{1}{4}\Big((\mathbb{I}+\vec{\tau}\cdot \vec{\sigma})\otimes 
\mathbb{I}-\vec{\lambda}\cdot(\vec{\sigma}\otimes \vec{\sigma})\Big).
\end{equation}
The positivity of the non-unital channel is guaranteed by $\rho_{AB}\geq 0$. Let us define $\tau=\parallel\vec{\tau}\parallel$ and 
$\hat{n}=\frac{\vec{\tau}}{\tau}$. Then the non-unital map is positive iff
\begin{eqnarray}
 q_{ij}\geq 0\: \:\mbox{and} \:\:\tau^2\leq u-\sqrt{u^2-q},
\end{eqnarray}
where $u=1-\sum \lambda_i^2+2\sum \lambda_i^2n_i^2$ and $q=\prod q_{ij}$ \cite{cohereB8}.

The coherence of the non-unital channel is given by 
\begin{equation}
C_{l_1}=\frac{1}{2}\Big(|\lambda_x+\lambda_y|+|\lambda_x-\lambda_y|+2\sqrt{\tau_x^2+\tau_y^2}\Big).
\label{nu-coh}
\end{equation}
Note that the coherence is independent of both $\lambda_z$ and $\tau_z$. Hence, some isocoherence planes will lie 
on the $\lambda_z$ and $\tau_z$ planes.
The purity for the channel is given by
\begin{equation}\label{nonunital purity}
\mathcal{P}=\frac{1}{4}(1+|\vec{\lambda}|^2+|\vec{\tau}|^2).
\end{equation}
Eq. (\ref{nonunital purity}) can be written as $|\vec{\lambda}|^2+|\vec{\tau}|^2=4\mathcal{P}-1$. 
It is clear, as before, that for a fixed purity, the values of parameters characterizing non-unital channels 
lie on the surface of a sphere. 
By fixing purity we can get the allowed regions of coherence as is shown in the Fig.\ref{unital-nonunital-reg}.

\noindent\textbf{Observation 1:} If the coherence of the channel is more than $1$, then it is non unital. 
One can easily see this from Fig. \ref{unital-nonunital-reg}. Hence, CoPu diagrams can help us distinguishing between unital and non unital channels for some region.

\noindent \textbf{Observation 2:} Unital channel cannot create coherence in the subsystem $A$ whereas the non-unital channel can. 

It can be easily checked by looking at the density matrix of the subsystem $A$  after the operation of the channel on the state (\ref{singlet state}). For unital and non-unital channel density matrices of subsystem $A$ are respectively
\begin{equation}
\rho_A^u=\frac{1}{2}\mathbb{I}
\quad
\mbox{and}
\quad
\rho_A^{nu}=\frac{1}{2}\begin{pmatrix}
1+\tau_z & \tau_x-i\tau_y \\
\tau_x+i\tau_y & 1-\tau_z
\end{pmatrix}.
\label{subcoh}
\end{equation}
where $\rho_A^u={\rm Tr}[\Phi_A^u\otimes \mathbb{I}_B(\ket{\Psi}_{AB}\bra{\Psi})]$ and 
$\rho_A^{nu}={\rm Tr}[\Phi_A^{nu}\otimes \mathbb{I}_B(\ket{\Psi}_{AB}\bra{\Psi})]$. 
Note that the non-unital channels with $\vec{\tau}=(0,0,\tau_z)$, can not create coherence in subsystem $A$.

\begin{figure}[h]
\centering
\includegraphics[scale=0.5]{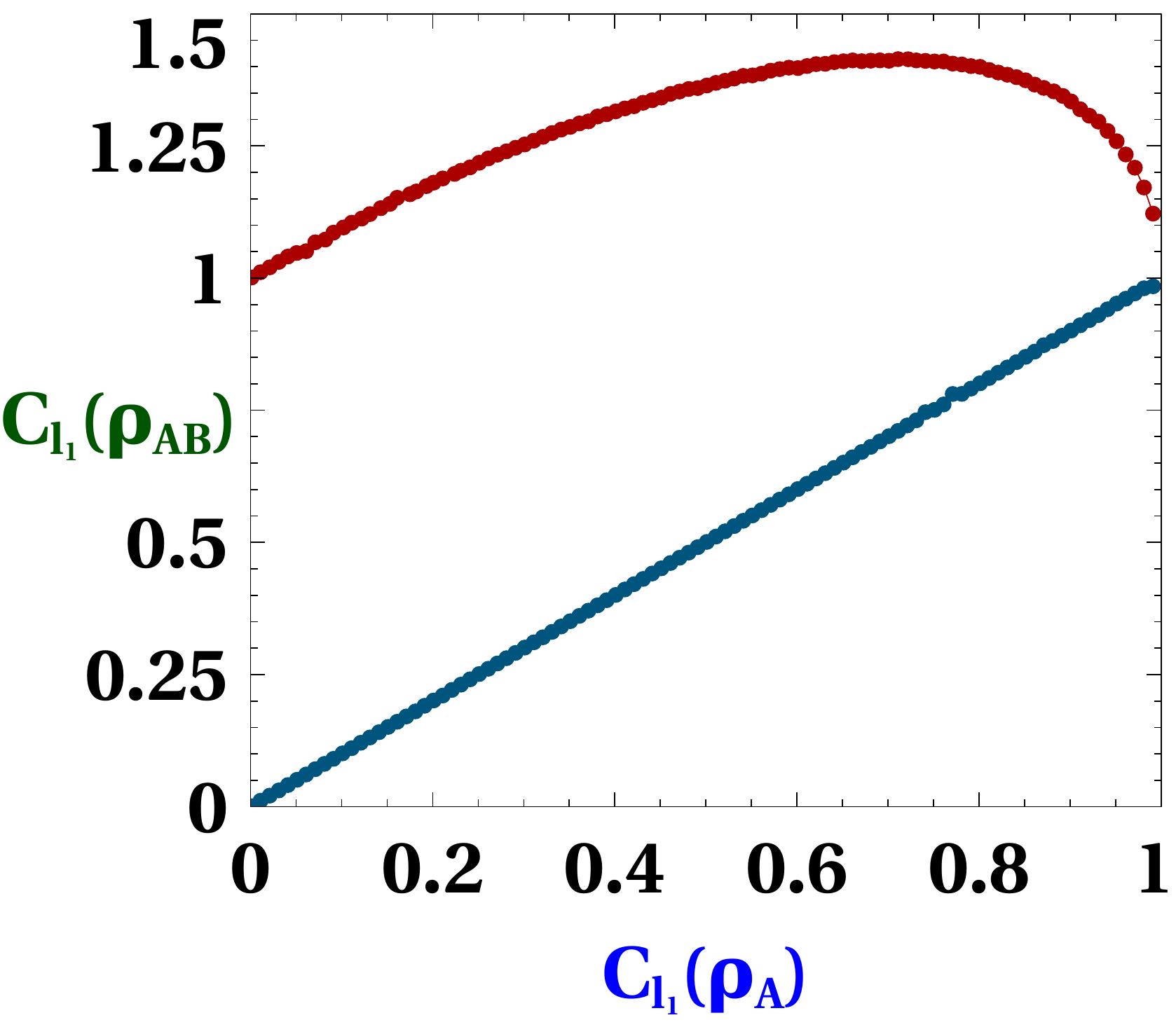}
\caption{(Color online) The channel coherence ($C_{l_1}(\rho_{AB})$) vs coherence induced in the subsystem $A$ ($C_{l_1}(\rho_{A})$) plot for nonunital qubit channels. The red curve depicts the nonunital channels which has maximum coherence for a given subsystem coherence whereas blue one represents the nonunital channels with minimum coherence. Plot shows that the minimum channel coherence 
is exactly equal to the coherence induced in the subsystem $A$.}
\label{nu_sub_coh}
\end{figure}

If we look closely at Eq.(\ref{nu-coh}) and Eq.(\ref{subcoh}), it is clear that the coherence of the nonunital channel can exactly 
be decomposed into the coherence of unital channel plus the coherence induced in the subsystem $A$ by the 
nonunital channel, i.e., 
\begin{equation}
 C_{l_1}(\Phi^{nu})=C_{l_1}(\Phi^{u})+C_{l_1}(\rho_{A}^{nu}).
 \label{coh-decomp}
\end{equation}
Also the Eq.(\ref{coh-decomp}) tells that $C_{l_1}(\Phi^{nu})\geq C_{l_1}(\Phi^{u})$. The Eq.(\ref{coh-decomp}) has been depicted in Fig.\ref{nu_sub_coh}. Plot shows that the minimum coherence of nonunital channels 
are exactly equal to the coherence induced in the subsystem $A$.

\noindent\textbf{Proposition 1:} All coherence breaking channels have zero coherence.\\
\textit{Proof.} A quantum channel is called coherence breaking channel if it maps any state to an incoherent state \cite{cbc}. This fact directly imply the above proposition.1. As an example, one 
can cosider the case of qubit channels. For coherence breaking qubit channels, the $\Lambda_{\Phi}$ should take the following form \cite{cbc}
\begin{equation}
\Lambda_{\Phi}=\begin{pmatrix}
1 & 0 & 0 & 0 \\
0 & 0 & 0 & 0 \\
0 & 0 & 0 & 0 \\
\tau_z & 0 & 0 & \lambda_z \\
\end{pmatrix}.
\nonumber
\end{equation}
Applying this channel on the state (\ref{singlet state}), one can show that Choi matrix is
\begin{equation}\label{choi coherence breaking}
\frac{1}{4}\begin{pmatrix}
1+\tau_z-\lambda_z & 0 & 0 & 0 \\
 0 & 1+\tau_z+\lambda_z & 0 & 0 \\
 0 & 0 & 1-\tau_z+\lambda_z & 0 \\
 0 & 0 & 0 & 1-\tau_z-\lambda_z \\
\end{pmatrix}.\nonumber
\end{equation}
Clearly, we get an incoherent Choi matrix.

%
However, note that this may not be the case for all entanglement breaking channels. It may happen that the 
final Choi matrix is separable but has coherence in subsystems.
 
\section{Coherence of incoherent channels}
The concept of incoherent operations is not unique. There exist many classes of incoherent operations
 in the literature. Below we consider the most important classes of incoherent operations that have been
discussed in the different resource theoretic perspectives of the quantum coherence.

\textit{Maximally Incoherent operation (MIO) :} A channel $\Phi$ is MIO iff $\Phi[\delta]\in \mathcal{I}$, 
for all incoherent states 
$\delta$, i.e., MIO preserves the set of incoherent states. This is the largest set of operations which 
preserve incoherence \cite{mio-aberg}.

\textit{Incoherent operation (IO) :} In Ref. \cite{cohereC4}, a smaller and relevant class of incoherent 
operations was introduced. A channel $\Phi$ with Kraus decomposition $\{K_i\}$ is IO iff 
$\frac{K_i \delta K_i^{\dag}}{\Tr[K_i \delta K_i^{\dag}]} \in \mathcal{I}$ for all $i$ and $\delta\in \mathcal{I}$. Hence, the Kraus operators of IO may be expressed as 
\begin{equation}
 K_i=\sum_{j=0}^{d-1} c_{ij}\ketbra{f_i(j)}{j},
 \label{IO_form1}
\end{equation}
where $f_i: \{0,1,..,d-1\}\mapsto \{0,1,..,d-1\} $ and $d$ is the dimension of the Hilbert space.
Note that coherence cannot be generated, even probabilistically, from incoherent states due to the action of this channel. 

The above two incoherent operations are defined in terms of their inability to create coherence. One can add 
further desirable restriction to the set of free operations. One such constraint is that the operations will be 
unable to use the coherence of the input state.

\textit{Strictly Incoherent operation (SIO) :} A channel $\Phi$ is SIO iff its Kraus operators $\{K_i\}$ individually 
commutes with dephasing, i.e., $\triangle (K_i\delta K_i^{\dag})=K_i\triangle (\delta) K_i^{\dag}$, 
where $\triangle$ is dephasing operation \cite{cohereR1,cohereC5} defined as 
$\triangle (\rho)=\sum_i\bra{i}\rho\ket{i}\ketbra{i}{i}$. This condition makes $f_i$ one-to-one, i.e., 
$f_i$ becomes permutation, $\pi_i$ in Eq.(\ref{IO_form1}).  Thus, SIO admits the set  
$\{K_i\}$ as well 
as $K_i^{\dag}$ are also incoherent. This indicates that the SIOs are not capable of using coherence of 
initial input states \cite{cohereC5}.

The above mentioned operations cannot be implemented by introducing an incoherent environment and a global 
unitary operation. This observation led one to introduce physically motivated incoherent operations 
\cite{cohereC9, pio-eric}.

\textit{Physical Incoherent operation (PIO) :} PIO is obtained through a class of noncoherence generating operations on a primary ($A$) and an ancillary system ($B$) \cite{cohereC9, pio-eric}.  A general PIO operation consist of an unitary operation $U_{AB}$ on the state $\rho_{A}$ of system $A$ and the incoherent state $\rho_B$ of system B, followed by a general incoherent projective measurement on system $B$. The PIO admits following Kraus decomposition 
$K_i=\sum_j e^{\mi\theta_j}\ketbra{\pi_i(j)}{j}P_i$ and their convex combinations. The $\pi_i$ are 
permutations and $\{P_i\}$ is an complete set of orthogonal incoherent projectors\cite{cohereC9}. Orthogonal incoherent projectors are those which does not introduce any coherence in the system after measurement. 

The PIOs are implementable 
using the aforementioned method and additionally it allows incoherent measurements in environment and classical 
post-selection on the outcomes. 

It is evident now that the MIO is the largest set of incoherent operations, and others are strict subset of it. 
The nontrivial relationship can be depicted in the following way \cite{cohereC9, pio-eric,pio-eric1,pio-eric2}
\begin{equation}
 PIO\subset SIO \subset IO \subset MIO.
\end{equation}
A special subset of PIO is considered and discussed in \cite{cohereA1}. These are very important in the sense that 
they preserve coherence of the input states.

\textit{Coherence preserving operation (CPO) :} A channel $\Phi$ is CPO iff it keeps the 
coherence of a state invariant, i.e., $C(\Phi[\rho])=C(\rho)$, where $C$ is an arbitrary 
coherence measure. The Kraus operator of CPO is expressed as $K=\sum_i e^{\mi\theta_i}\ketbra{\pi(i)}{i}$.

We also consider the following two incoherent operations introduced in Ref.\cite{pio-eric2}.

\textit{Genuinely Incoherent operation (GIO) :} A channel $\Phi$ is GIO iff $\Phi[\delta]=\delta$, i.e., 
all incoherent states are fixed points for the channel. Therefore, GIO does not allow transformation 
between any incoherent states. 
All Kraus operators for this operation are diagonal in the incoherent basis.

\textit{Fully Incoherent operation (FIO) :} A quantum operation is fully incoherent if and only if all 
Kraus operators are incoherent and have the same form. Kraus operators are incoherent means, 
$K_i \delta K^\dagger_i$ is an incoherent state as well. This means that only pure incoherent states 
are free in this resource theory.
%

There exist other concepts of free operations in the context of resource theory of coherence \cite{cohereR}, but 
we limit ourselves to the operations described above.

The following Kraus representations are the possible FIOs for single qubits 
\begin{eqnarray}\label{FIO}
\left\{ \begin{pmatrix}a_1 & b_1\\
0 & 0
\end{pmatrix},\begin{pmatrix}a_2 & b_2\\
0 & 0
\end{pmatrix}\right\};
\left\{\begin{pmatrix}0 & 0\\
a_1 & b_1
\end{pmatrix},\begin{pmatrix}0 & 0\\
a_2 & b_2
\end{pmatrix}\right\};\nonumber\\
\left\{ \begin{pmatrix}0 & d_1\\
c_1 & 0
\end{pmatrix},\begin{pmatrix}0 & d_2\\
c_2 & 0
\end{pmatrix}\right\};
\left\{\begin{pmatrix}c_1 & 0\\
0 & d_1
\end{pmatrix},\begin{pmatrix}c_2 & 0\\
0 &  d_2
\end{pmatrix}\right\};
\end{eqnarray}
where $|a_1|^2+|b_1|^2=1=|a_2|^2+|b_2|^2$ and $a_1b_1^*+a_2b_2^*=0=b_1a_1^*+b_2a_2^*$, and $|c_1|^2+|c_2|^2=1=
|d_1|^2+|d_2|^2$. From Eq. (\ref{FIO}) one can easily check that all the matrices of a Kraus representations have the same form. As an example, the first two matrices have nonzero entries only in the first row. The last one is the GIO for the qubit case. Note that first two FIOs have zero coherence. The coherence 
and purity of last two FIOs are $C_{l_1}=|d_1c_1^*+d_2c_2^*|$ and $\mathcal{P}=\frac{1}{2}(1+C_{l_1}^2)$. 
Hence, we have the relation $2\mathcal{P} - C_{l_1}^2=1$ with $\mathcal{P}\in[\frac{1}{2},1]$.

According to the Ref. \cite{forms-of-ios}, any qubit incoherent operation (IO) admits a decomposition with at most five 
Kraus operators. A canonical choice of Kraus operators for IO is
\begin{equation}
\left\{ \begin{pmatrix}a_{1} & b_{1}\\
0 & 0
\end{pmatrix},\begin{pmatrix}0 & 0\\
a_{2} & b_{2}
\end{pmatrix},\begin{pmatrix}a_{3} & 0\\
0 & b_{3}
\end{pmatrix},\begin{pmatrix}0 & b_{4}\\
a_{4} & 0
\end{pmatrix},\begin{pmatrix}a_{5} & 0\\
0 & 0
\end{pmatrix}\right\} ,\label{eq:IO-qubit}
\end{equation}
where one can choose $a_i\in\mathbb{R}$ while $b_{i}\in\mathbb{C}$. Further, 
$\sum_{i=1}^{5}a_{i}^{2}=\sum_{j=1}^{4}|b_{j}|^{2}=1$
and $a_{1}b_{1}+a_{2}b_{2}=0$ holds. The coherence and purity of 
IOs are $C_{l_1}=\sum_{i=1}^4 a_i|b_i|$ and $\mathcal{P}=\frac{1}{2}[1-\mu(1-\mu)-\kappa(1-\kappa)+\sum_{i=1}^4 a_i^2|b_i|^2]$, respectively, 
where $\mu=(a_2^2+a_4^2)$ and $\kappa=(|b_1|^2+|b_4|^2)$. 

Similarly, Ref. \cite{forms-of-ios} shows that the canonical set of Kraus operators for SIO is
\begin{equation}
\left\{ \begin{pmatrix}a_{1} & 0\\
0 & b_{1}
\end{pmatrix},\begin{pmatrix}0 & b_{2}\\
a_{2} & 0
\end{pmatrix},\begin{pmatrix}a_{3} & 0\\
0 & 0
\end{pmatrix},\begin{pmatrix}0 & 0\\
a_{4} & 0
\end{pmatrix}\right\} ,
\end{equation}
where $a_i\in\mathbb{R}$ and 
$\sum_{i=1}^{4}a_{i}^{2}=\sum_{j=1}^{2}|b_{j}|^{2}=1$ holds. The coherence and purity of 
SIOs are $C_{l_1}=a_1|b_1|+a_2|b_2|$ and $\mathcal{P}=\frac{1}{2}[1-\nu(1-\nu)+|b_1|^2|b_2|^2+\sum_{i=1}^2 a_i^2|b_i|^2]$, respectively, 
with $\nu=(a_1^2+a_3^2)$. 

\begin{figure}[h]
\centering
\includegraphics[scale=0.5]{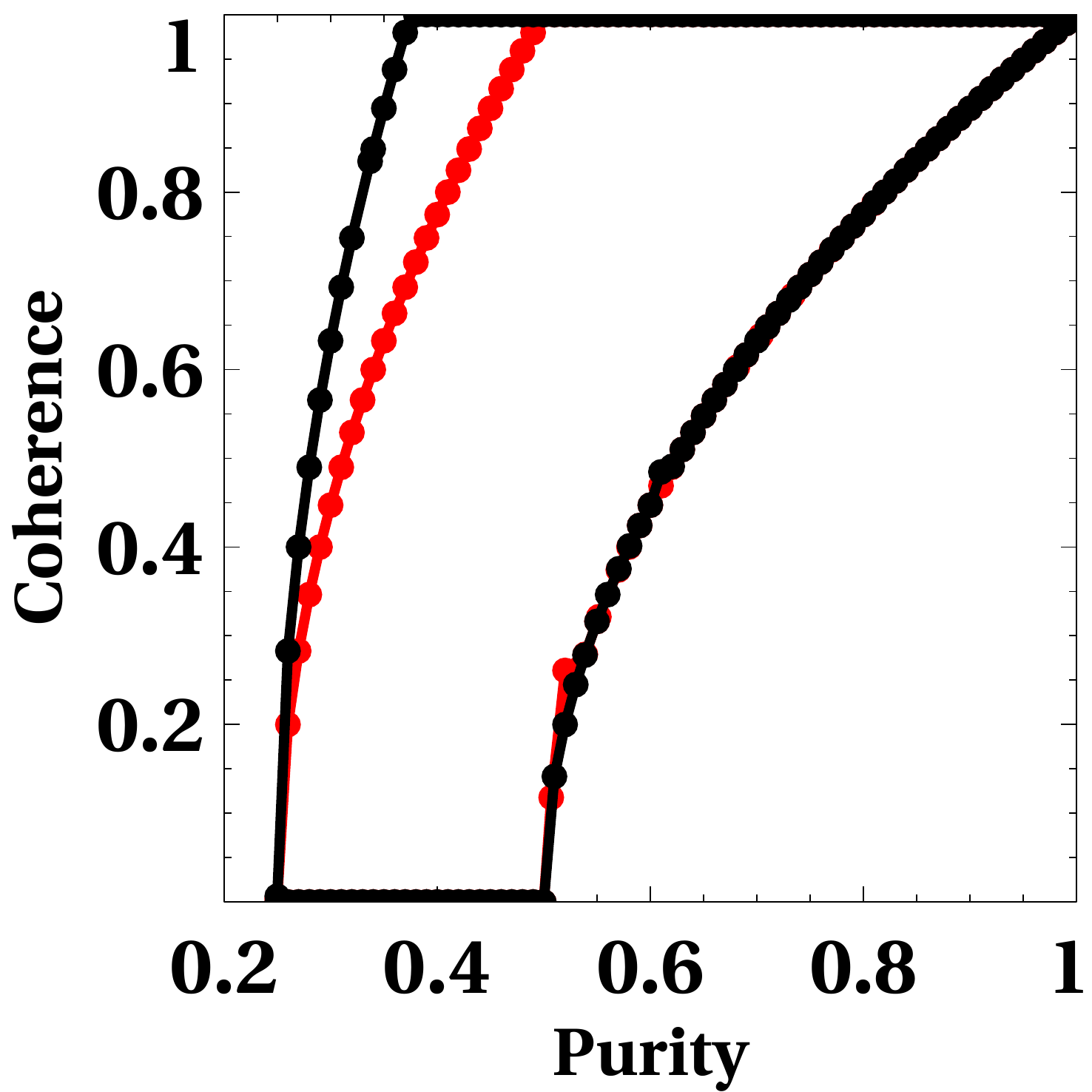}
\caption{(Color online) The allowed coherence-vs-purity region for IO (region inside black curve) and SIO (region inside red curve) respectively. Coherence of the channels is measured by $l_1$-norm. The figure depicts the well known phenomenon that $SIO \subset IO$. Moreover, channels outside the overlap region are IO and can be easily distinguished from the SIO.}
\label{IO-Sio-copu}
\end{figure}

The CoPu diagrams in Fig.(\ref{IO-Sio-copu}) show that SIOs are subset of IOs. Note that all of the purity range is not allowed for both SIOs and IOs.

\noindent\textbf{Observtion 3:} It is possible to distinguish between SIO and IO for some regions of CoPu diagram in  Fig. \ref{IO-Sio-copu}. According to the Ref.\cite{forms-of-pios, forms-of-ios}, if one considers state transformation by incoherent 
operations, qubit SIOs and IOs are equivalent. However, the above observation tells us the opposite behavior.

The Kraus representation of all possible single qubit PIOs are given by \cite{forms-of-pios}
\begin{eqnarray}
\left\{ \begin{pmatrix}e^{i\theta_1} & 0\\
0 & 0
\end{pmatrix},\begin{pmatrix}0 & 0\\
0 & e^{i\theta_2}
\end{pmatrix}\right\};
\left\{\begin{pmatrix}0 & 0\\
e^{i\phi_2} & 0
\end{pmatrix},\begin{pmatrix}0 & e^{i\phi_1}\\
0 & 0
\end{pmatrix}\right\};\nonumber\\
\left\{ \begin{pmatrix}e^{i\theta_1} & 0\\
0 & 0
\end{pmatrix},\begin{pmatrix}0 & e^{i\phi_1}\\
0 & 0
\end{pmatrix}\right\};
\left\{\begin{pmatrix}0 & 0\\
e^{i\phi_2} & 0
\end{pmatrix},\begin{pmatrix}0 & 0\\
0 &  e^{i\theta_2}
\end{pmatrix}\right\};\\
\left\{ \begin{pmatrix}e^{i\theta_1} & 0\\
0 & e^{i\theta_2}
\end{pmatrix}\right\};\left\{\begin{pmatrix}0 & e^{i\phi_1}\\
e^{i\phi_2} & 0
\end{pmatrix}\right\},~~~~~~~~~~~~~~~~~
\end{eqnarray}
where first four PIOs are the coherence breaking channels and have zero coherence in both KROC and AROC, 
and the last two PIOs are the all possible single qubit CPOs and have unit coherence and unit purity in KROC.

Although we have expected that the coherence of the incoherent channels will be zero, it turns out to be 
not so. However, we draw the following observation from the incoherent channels considered in this section.  

\noindent\textbf{Observation 4:} All qubit incoherent channels which are either unital or nonunital, 
cannot create coherence in the subsystem `$A$' of its Choi matrix. It can easily be verified from the 
Table.I.

\begin{proof}
 Here we will try to prove the Observation.4 for IO, SIO and PIO. If we consider the 
 Kraus decomposition of IO as given in Eq.(\ref{IO_form1}), then its Choi matrix will be 
 \begin{eqnarray}
  \rho_{AB}&=&\frac{1}{d}\sum_{i}K_i\otimes\mathbb{I}\left(\sum_{lm}\ketbra{ll}{mm}\right)K_i^\dagger\otimes\mathbb{I},\nonumber\\
  &=&\frac{1}{d}\sum c_{ij}c_{it}^*\ketbra{f_i(j)}{j}l\rangle\langle m\ketbra{t}{f_i(t)}\otimes \ketbra{l}{m},\nonumber\\
  &=& \frac{1}{d}\sum c_{ij}c_{it}^*\ketbra{f_i(j)}{f_i(t)}\otimes \ketbra{l}{m} \delta_{jl}\delta_{mt},\nonumber\\
  &=& \frac{1}{d}\sum c_{il}c_{it}^*\ketbra{f_i(l)l}{f_i(t)t}.\nonumber
 \end{eqnarray}
Now the reduced density matrix of the subsystem $A$ is 
\begin{eqnarray}
 \rho_A &=&\frac{1}{d}\sum c_{il}c_{it}^*\ketbra{f_i(l)}{f_i(t)}\otimes \langle n\ketbra{l}{t}n\rangle,\nonumber\\
 &=& \frac{1}{d} \sum c_{il}c_{il}^*\ketbra{f_i(l)}{f_i(l)}.
\end{eqnarray}
Therefore, $\rho_A$ is incoherent for IO. This also guarantees that $\rho_A$ will be incoherent for SIO and 
PIO. Although we do not have direct proof for other type of incoherent operations, the following Table.I confirms 
that the Observation.4 is also true atleast for single qubit FIO and GIO.
\end{proof}

This observation says that the nonunital channels which has $\vec{\tau}=\{0,0,\tau_z\}$ qualifies as 
potential candidates for incoherent operations (see Table.I).

\begin{table}[h]
\label{inc-tab}
\begin{center}
\begin{tabular}{c c c c}
\hline
\multirow{2}{1.5cm} & \multicolumn{2}{c}{Coherence of} \\
\cline{2-3}
Channels    & $\rho_{AB}$ & $\rho_A$ & $~~~$ $\tau_z$\\
\hline
IO      & $[0,1]$    & 0     & $~~~$*  \\
SIO          &  $[0,1]$       & 0      & $~~~$*  \\
PIO (CPO)      & 0 (1)    & 0      & $~~~$* \\
FIO (GIO) & \#($[0,1]$)      & 0       & $~~~~~$\# (0)\\
\hline
\end{tabular}
\end{center}
\caption{Table shows that all qubit incoherent operations have zero coherence in $\rho_A(={\rm Tr}_A[\rho_{AB}]$). The $*$ denotes that  
the corresponding channels are in general nonunital. The \# for FIOs indicates 
that the channels which have zero coherence (in Choi matrix) are nonunital otherwise they are unital.}
\end{table}

\subsection{Coherence Non-Generating Channel (CNC)}
A CPTP map, $\Phi$ which does not generate quantum coherence from an incoherent state is known as 
the coherence non-generating channel \cite{Cngc}, i.e.,  $\Phi[\mathcal{I}]\subset\mathcal{I}$. The incoherent 
operations are strict subset of these channels. These channels are different from the set of incoherent operations 
in the sense that the monotonicity of coherence may break under these operations while acting on one subsystem \cite{Cngc}.

\noindent\textbf{Proposition 2:} For general qubit CNC channels, $0\leq C_{l_1}\leq\sqrt{2}$.\\
\textit{Proof.} A full rank qubit channel is CNC iff it admits following two Kraus decompositions \cite{Cngc}. The first one is 
$$K_1=\begin{pmatrix}
\me^{\mi\eta}\cos\theta\cos\phi & 0 \\
-\sin\theta\sin\phi & \me^{\mi\xi}\cos\phi
\end{pmatrix},$$ 

$$ K_2=\begin{pmatrix}
\sin\theta\cos\phi & \me^{\mi\xi}\sin\phi  \\
\me^{-\mi\eta}\cos\theta\sin\phi & 0
\end{pmatrix},$$
where $\theta,\phi,\xi,\eta\in \mathbb{R}$. Notice that $K_1$ and $K_2$ may not individually be incoherent but $K_1(\cdot)K_1^\dagger+K_2(\cdot)K_2^\dagger$ 
can be 
if $\sin\phi\cos\phi\sin\theta\cos\theta=0$. Therefore, CNC channels may not be incoherent. 

The coherence and purity of the above channel are given by  
$C_{l_1}=\cos\theta+|\sin\theta\sin 2\phi|$ and $\mathcal{P}=\frac{1}{8}(5+\cos 2\theta+2\cos^2\theta\cos 4\phi)$, 
respectively. The incoherent condition will always guarantee that the 
coherence will be less than or equal to $1$. Otherwise, the coherence of CNC can be $0\leq C_{l_1}\leq \sqrt{2}$. 
The coherence will reach its maximum at $\theta=\frac{\pi}{4}=\phi$.

The other CNC channel is given by
$$K_1=\begin{pmatrix}\cos\theta & 0 \\
0 & \me^{\mi\chi}\cos\phi
\end{pmatrix}\quad\mbox{and}\quad K_2=\begin{pmatrix}
0 & \sin\phi \\
\me^{\mi\chi}\sin\theta & 0
\end{pmatrix}.$$
This channel is an incoherent channel. The coherence and purity of this CNC 
channel is $C_{l_1}=\cos\theta\cos\phi+|\sin\theta\sin\phi|$ and $\mathcal{P}=\frac{1}{16}(10+\cos 4 
\theta+4\cos 2\theta\cos 2 \phi+\cos 4\phi)$, respectively and $0\leq C_{l_1}\leq1$.

\begin{figure}[h]
\centering
\includegraphics[scale=0.5]{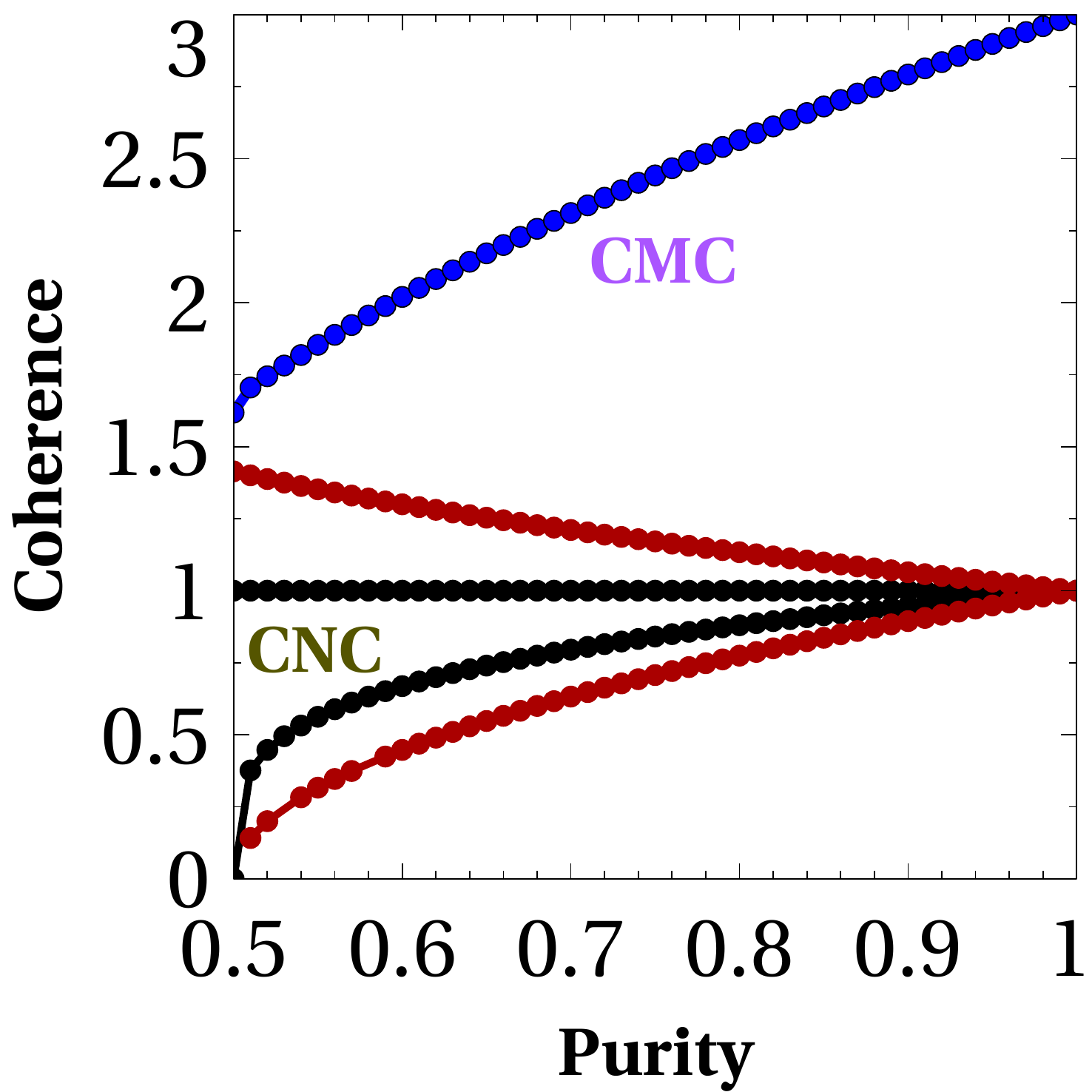}
\caption{(Color online) The allowed coherence-vs-purity region for CMC and CNC. 
The region inside the red curve is for CNC and if the CNCs are incoherent then its coherence lie in 
the region between black curves. The region between upper black line ($y=1$ -line) and blue curves is for CMCs. Note that 
for all these channels $\mathcal{P}\in[\frac{1}{2},1]$.}
\label{cnc-cmc}
\end{figure}

The Fig. (\ref{cnc-cmc}) shows allowed range of all CNC channels. It is clear that allowed region of incoherent CNCs is inside 
the region of all CNCs. From the CoPu diagrams it is clear that for these channels, purity ranges from $\frac{1}{2}$ to $1$.  

\section{Coherence of other known qubit channels}
In this section we will consider some known qubit channels and find its $l_1$-norm coherence. We 
will investigate whether these channels can be characterized by its coherence and purity.\\
\textit{\textbf{Channel to obtain maximum coherence (CMC):}} The maximum value of the $l_1$-norm coherence for 
two qubit system is $3$. This value is achieved by the state $\ket{++}$, where 
$\ket{+}=\frac{1}{\sqrt{2}}(\ket{0}+\ket{1})$. Hence, its obvious to 
search for a qubit channel which will reach this value.

\noindent\textbf{Proposition 3:} For general two qubit CMC channels, $1\leqslant C_{l_1}\leqslant 3$.
\\
\textit{Proof.} The channel which may reach this value admits 
the following Kraus decomposition 
$$K_1=\frac{1}{\sqrt{2}}\begin{pmatrix}
\cos\theta_1 & \me^{-\mi\phi_1}\sin\theta_1 \\
\me^{\mi\phi_1}\sin\theta_1 & -\cos\theta_1
\end{pmatrix},$$ 

$$K_2=\frac{1}{\sqrt{2}}\begin{pmatrix}
\cos\theta_2 & \me^{-\mi\phi_2}\sin\theta_2 \\
\me^{\mi\phi_2}\sin\theta_2 & -\cos\theta_2
\end{pmatrix}.$$ 
The coherence and purity of the channel is 
\begin{eqnarray}
 C_{l_1}&=&\frac{1}{4}\Big(2+\varsigma+\sum_{j=\pm}g_j+f_j\Big),\nonumber\\
 \mathcal{P}&=&\frac{1}{16}(11+3\cos 2\theta_1\cos 2\theta_2+\varsigma+\ell_{21}+\ell_{12}),
\end{eqnarray}
where $g_{\pm}=|\me^{\pm 2\mi\phi_1}\sin^2\theta_1+\me^{\pm 2\mi\phi_2}\sin^2\theta_2|$, $f_{\pm}=2|\me^{\pm \mi\phi_1}\sin2\theta_1+\me^{\pm\mi\phi_2}\sin2\theta_2|$, 
$\varsigma=\cos2\theta_1+\cos2\theta_2$ and $\ell_{mn}=4\cos m(\phi_1-\phi_1)\sin^m n\theta_1\sin^m n\theta_2$. 
The coherence of the channel will reach its maximum, i.e., 
$C_{l_1}=3$ for $\theta_1=\frac{\pi}{4}=\theta_2$ and $\phi_1=\phi_2$. In fact, the coherence of this channel obeys 
$1\leq C_{l_1}\leq 3$ which is confirmed by the CoPu diagram (see Fig. (\ref{cnc-cmc})). Therefore, CMC channel either increases or unalters the coherence of the state. Moreover, this channel can be considered as coherence generating channel.  
\begin{figure}[h]
\centering
\includegraphics[scale=.6]{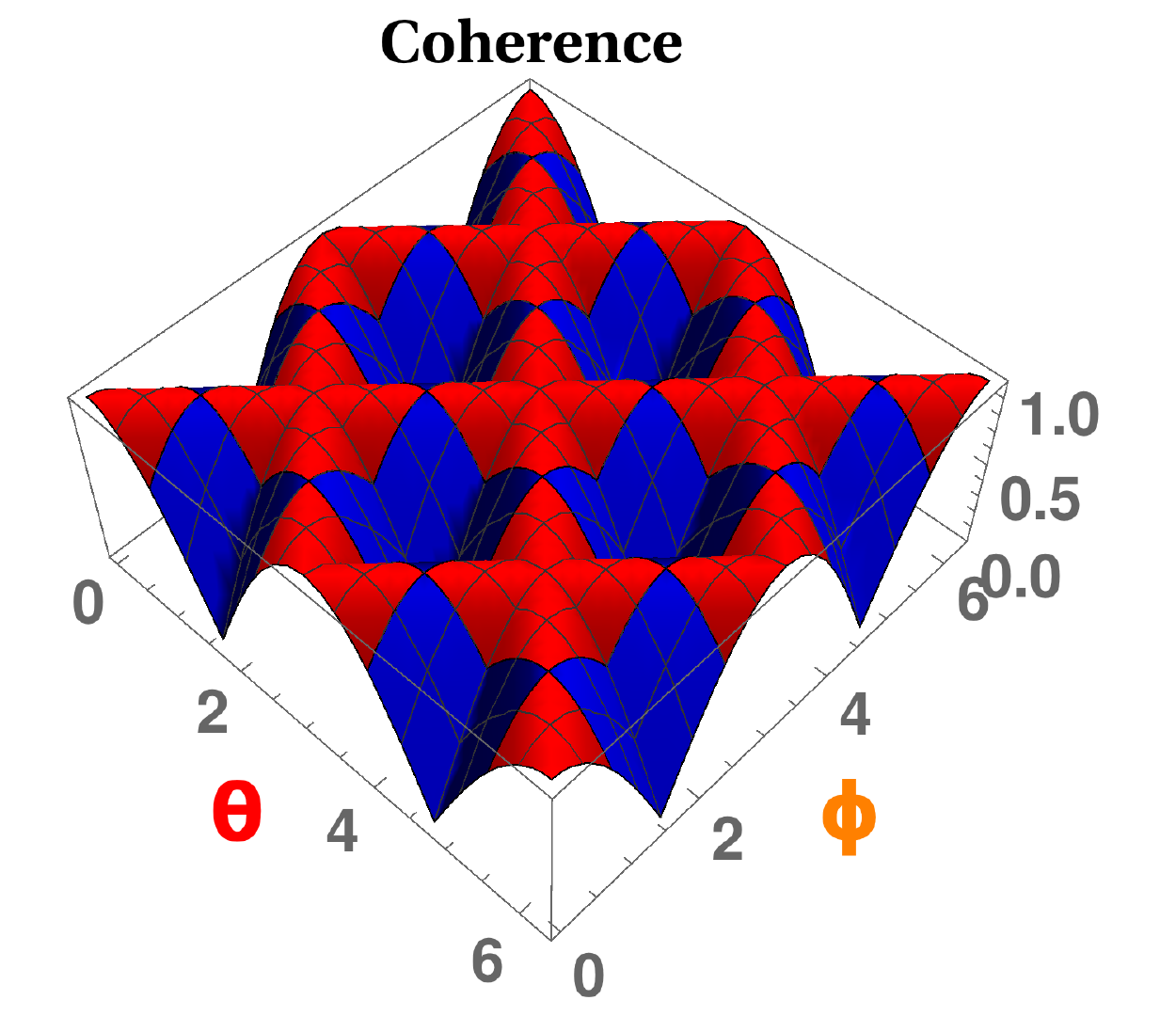}
\caption{(Color online) Plot of the $l_1$-norm coherence of degradable (red regions) and anti-degradable channels (blue regions) with the parameters $\theta$ and $\phi$. 
It depicts that the whole region is completely covered and degradable channels lie in the range $\frac{1}{\sqrt{2}}\leq C_{l_1}\leq 1$.}
\label{deg-antd}
\end{figure}

\textit{\textbf{A family of qubit channels:}} 
A family of qubit channels can be described by two Kraus operators 
in $\sigma_z$ basis as \cite{afchannel}
\begin{equation}
K_1=\begin{pmatrix}\cos\theta & 0 \\
0 & \cos\phi
\end{pmatrix}\quad\mbox{and}\quad K_2=\begin{pmatrix}
0 & \sin\phi \\
\sin\theta & 0
\end{pmatrix},
\label{qubit_paul}
\end{equation}
where $\theta,\phi\in [0,\pi]$. In AROC, the channel is described by $\lambda_x=\cos(\phi-\theta)$, 
$\lambda_y=\cos(\phi+\theta)$, $\lambda_z=(\cos2\phi+\cos2\theta)/2$, 
$\tau_x=\tau_y=0$ and $\tau_z=(\cos2\theta-\cos2\phi)/2$.   
The coherence and purity of the channel are $C_{l_1}=\cos\theta\cos\phi+|\sin\theta\sin\phi|$ and $\mathcal{P}=\frac{1}{2}+\frac{1}{8}(\cos2\phi+\cos2\theta)^2$.

\begin{figure}[h]
\centering
\includegraphics[scale=0.55]{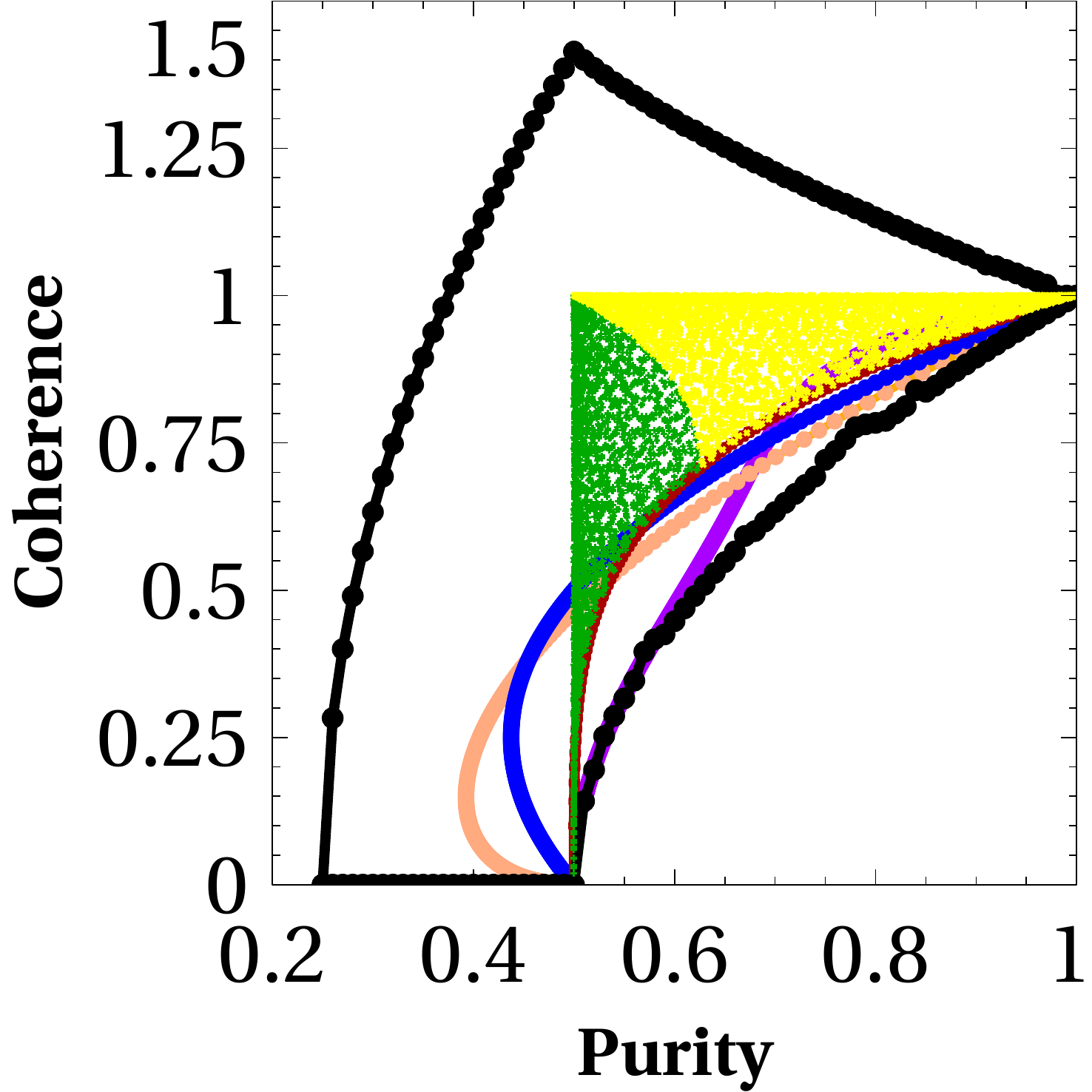}
\caption{(Color online) Figure depicts the coherence vs purity curve for non-unital channels. The green region represents the anti-degradable channels whereas 
the yellow region shows the degradable ones. The lower boundary of these two channels coincides with the red curve which represents the amplitude damping channels. 
The blue, orange, red and the purple curve represent the homogenization channels for $T_2=T_1$, $2T_2=T_1$, $T_2=2T_1$ and $T_2=5T_1$ respectively with $\omega=1$. 
Note that the $T_2=2T_1$ curve coincides with AD curve.}
\label{nu-all-cp}
\end{figure}

We know that a CPTP map can be described as a unitary coupling with the external environment. If a CPTP map $\Phi$ changes a state $\rho_S$ to $\rho_S^\prime$, then it can be represented as
\begin{equation}
\rho_S^\prime=\Phi(\rho_S)=\mbox{Tr}_E[U_{SE}(\rho_S\otimes \omega_E)U_{SE}^\dagger],
\end{equation}
where $U_{SE}$ is the unitary coupling between the system and the environment $E$ and $\omega_E$ is a fixed state of $E$. In this process environment state also changes to $\omega_E^\prime$. A CPTP map can also be represented as operator sum representation as in Eq. (\ref{kraus rep}). The final environment state can be found from the initial system state $\rho$ by using the complementary channel $\Phi^\prime$ of $\Phi$ as 
\begin{equation}
\omega_E^\prime=\Phi^\prime(\rho_S)=\mbox{Tr}_S[U_{SE}(\rho_S\otimes \omega_E)U_{SE}^\dagger].
\end{equation}
A map $\Phi$ is called degradable if there exits a third map $\Omega$ such that $\Phi^\prime=\Omega \Phi$ and $\Omega$ takes the state $\rho_S^\prime$ to $\omega^\prime_E$\cite{afchannel}. Similarly a map is antidegradable if there exist a $\Omega$ such that $\Phi=\Omega \Phi^\prime$ and which take the final environment state $\omega_E^\prime$ to $\rho_S^\prime$ \cite{afchannel}. More details can be found in  the reference \cite{afchannel}. The channel represented in Eq. (\ref{qubit_paul})  is degradable for $\frac{\cos2\theta}{\cos2\phi}\geq 0$, otherwise anti-degradable \cite{afchannel}, see Fig.(\ref{deg-antd}). The Fig.(\ref{deg-antd}) 
shows that the coherence of degradable channel satisfies $\frac{1}{\sqrt{2}}\leq C_{l_1}\leq 1$. The CoPu diagram in Fig.(\ref{nu-all-cp}) also confirms this 
observation.

\noindent\textbf{Observation 5:} If the channel in Eq.(\ref{qubit_paul}) is anti-degradable then its 
coherence be always less than $\frac{1}{\sqrt{2}}$.

For $\cos 2\theta=1$ and $\cos 2\phi=2\eta-1$, it describes the amplitude damping (AD) channels with damping rate $\eta$. 
The coherence and purity of AD channels are $C_{l_1}=\sqrt{\eta}$ and $\mathcal{P}=\frac{1}{2}(1+C_{l_1}^4)$ respectively. 
Equivalently, $C_{l_1}=\sqrt[4]{2\mathcal{P}-1}$. As it is a non-unital channel, the CoPu diagram for this channel is shown in Fig.(\ref{nu-all-cp}).

\begin{figure}[h]
\centering
\includegraphics[scale=0.5]{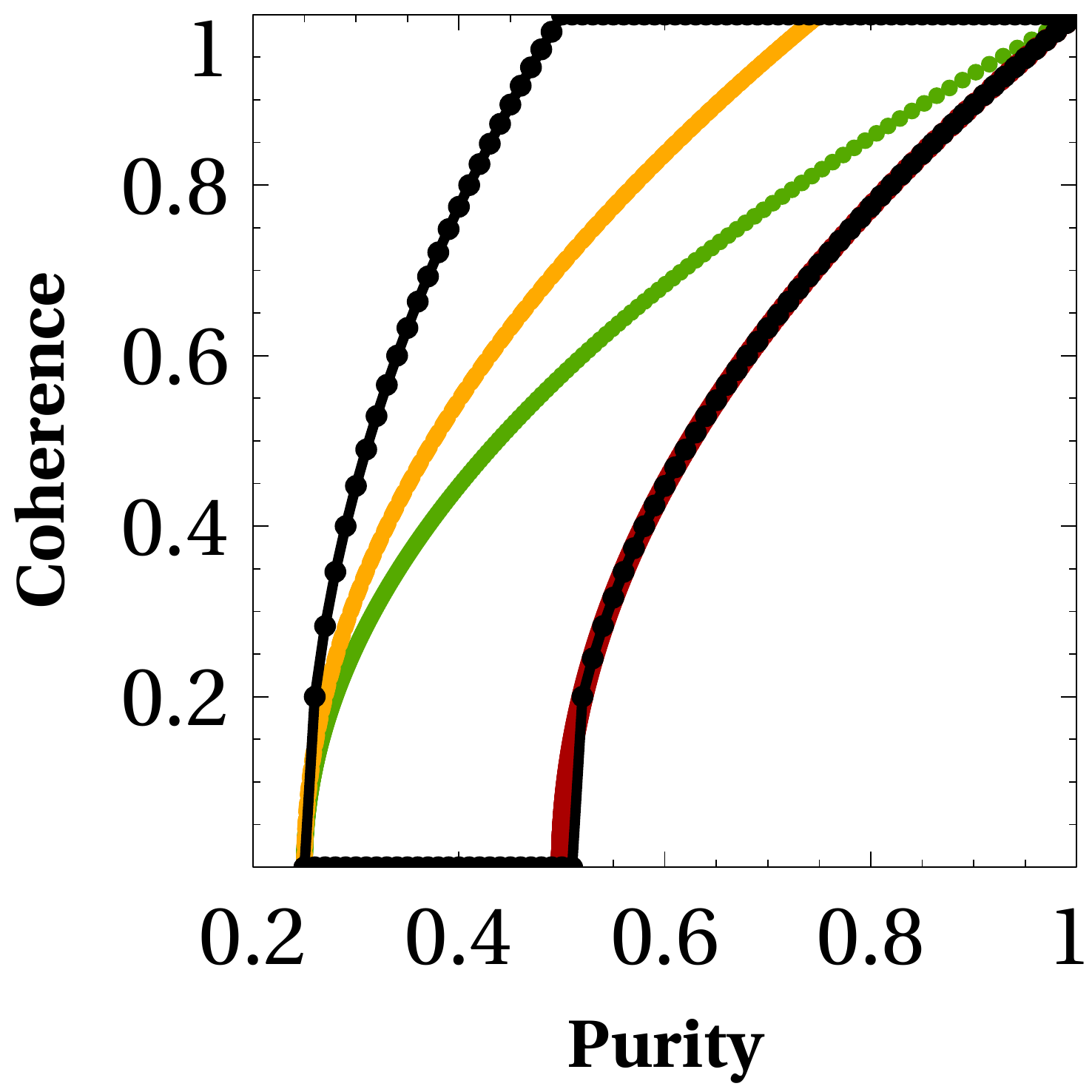}
\caption{(Color online) Figure depicts the coherence vs purity curve for unital channels. The red curve depicts the decoherence channels which coincides with the 
lower boundary of CoPu curve of unital channels. The green curve represents the depolarizing channels and the orange one is for the bit flip as well as bit-phase flip channels.}
\label{u-all-cp}
\end{figure}

If $\sin\theta=\pm \sin\phi$, the above channel becomes unital. Specifically, for $\theta=\phi$, the channel becomes a 
bit flip channel but for $\theta=-\phi$, it is a bit-phase flip channel. The coherence and purity of both bit flip and bit-phase 
flip channels are $C_{l_1}=\cos (2\theta)$ and $\mathcal{P}=\frac{1}{4}(1+2C_{l_1}^2)$ respectively or equivalently, 
$2\mathcal{P}-C_{l_1}^2=\frac{1}{2}$, with $\mathcal{P}\in [\frac{1}{4},\frac{3}{4}]$. 
The above channel will unitarily transform to 
a phase flip channel (decoherence channel) if we multiply the Kraus operators with Hadamard gate \cite{afchannel}. All the Pauli channels are unital channels. Thus,
the CoPu diagrams of these channels can be depicted inside the CoPu diagram of general unital channels (see Fig. (\ref{u-all-cp})).

\noindent\textit{\textbf{Qubit Decoherence, Depolarization and Homogenization channels:}} The decoherence, depolarization and 
homogenization are nonunitary channels and form Markovian semigroup \cite{cohereE2}. 

The decoherence is a process in which the off-diagonal terms of the density matrix of a quantum system are 
continuously suppressed in time, i.e., $\rho\rightarrow\rho_{t\rightarrow\infty}={\rm diag}(\rho)$. 
The decoherence channel is described by 
$\lambda_x=\lambda_y=\me^{-\frac{t}{T}}$ and $\lambda_z=1$. This channel is a unital channel. The coherence and purity of this channel are given by 
\begin{equation}
 C_{l_1}=\me^{-\frac{t}{T}} \:\:\mbox{and}\:\: \mathcal{P}=\frac{1}{2}(1+C_{l_1}^2)
\end{equation}
respectively. Now we have $2\mathcal{P}-C_{l_1}^2=1$. As $\mathcal{P}\in [\frac{1}{2},1]$, the decoherence channels will 
represent the minimum coherence boundary of unital channels. The CoPu diagram for this channel is in Fig.(\ref{u-all-cp}).

\noindent\textbf{Observation 6:} For the qubit decoherence channels, the concurrence and the $l_1$-norm coherence are same.\\
The above observation can be easily verified as the concurrence of the decoherence channel is $\me^{-\frac{t}{T}}$ \cite{cohereE2}.

The depolarizing channel with noise parameter $p$ transmits an input qubit perfectly with probability
$1-p$ and outputs the completely mixed state with probability $p$, i.e., $\rho\rightarrow\rho_f=(1-p)\rho + p\mathbb{I}/2$. The depolarization channel is described as $\lambda_x=\lambda_y=\lambda_z=\me^{-\frac{t}{T}}$. This channel is also a unital channel.
The coherence and purity of this channel are 
\begin{equation}
 C_{l_1}=\me^{-\frac{t}{T}} \:\:\mbox{and}\:\:  \mathcal{P}=\frac{1}{4}(1+3C_{l_1}^2), 
\end{equation}
respectively. Therefore, $4\mathcal{P}-3C_{l_1}^2=1$, with $\mathcal{P}\in [\frac{1}{4},1]$. This restriction 
is represented in CoPu diagram (see Fig.(\ref{u-all-cp})).

The homogenization is an evolution that transforms the whole Bloch sphere into a single point, i.e., 
it is a contractive map with the fixed point (the stationary state of the dynamics). This map is described by 
$\lambda_x=\lambda_y=\me^{-\frac{t}{T_2}}$, $\lambda_z=\me^{-\frac{t}{T_1}}$, 
$\tau_x=\tau_y=0$ and $\tau_z=\omega(1-\me^{-\frac{t}{T_1}})$, where the
parameters, $\omega$ is the purity of the final state, $T_1$ is the
decay time, $T_2$ is the decoherence time. It is a non-unital process. The coherence and the purity of this channel are 
\begin{equation}
 C_{l_1}=\me^{-\frac{t}{T_2}} 
 \:\:\mbox{and}\:\: \mathcal{P}=\frac{1}{4}[1+\me^{-\frac{2t}{T_1}}+2\me^{-\frac{2t}{T_2}}+\omega^2(1-\me^{-\frac{t}{T_1}})^2], 
\end{equation}
respectively. For $\omega=1$ and $T_2=2T_1$, we have the relation between the coherence and purity as given by  
$C_{l_1}=\sqrt[4]{2\mathcal{P}-1}$. Some CoPu diagrams of this channel are shown in Fig. (\ref{nu-all-cp}).

\section{Discussion and Conclusion}
The significance of this work is two fold: Choi-Jamio\l{}kowski isomorphism allows us to associate a  
density matrix with a channel. The purity and coherence of this density matrix can be fruitfully associated
with the channel. Using CoPu diagrams we have shown that it may be possible 
to distinguish different qubit channels.
Distinguishing the channels using CoPu diagrams depicts the inter-relation between purity and the 
coherence of the channels. It has a broader meaning also, e.g., given a purity one may not find a channel which has certain amount of coherence. These 
relations between coherence and purity show deeper restriction on available coherent channel.

If we look closely, we find that the purity and coherence satisfy the following 
equation in some of the cases, i.e., 
\begin{equation}
 \varpi \mathcal{P}-\varphi C_{l1}^2=1,
\end{equation}
where $\varpi,\varphi\in \mathbb{R}$. 

We can rewrite this as, 
\begin{equation}
 \left(\sqrt{\frac{\varpi}{\varphi}}\sqrt{\mathcal{P}}\right)^2-(C_{l1})^2=\left(\frac{1}{\sqrt{\varphi} }\right)^2.
 \label{4vec_pc}
\end{equation}

   This expression can have an interesting interpretation. We can think of scaled $\sqrt{\mathcal{P}}$ as
   time component, and coherence as the spatial component of a vector in two-dimensional Minkowski
   space. This vector is timelike.

Boundaries of light cone are given by the following relation
\begin{equation}
 \left(\frac{C_{l1}}{\sqrt{\mathcal{P}}}\right)^2=\frac{\varpi}{\varphi}.
\end{equation}
In general, $\frac{C_{l1}}{\sqrt{\mathcal{P}}}\leq\pm \sqrt{\frac{\varpi}{\varphi}}$. Physically, this means that 
the allowed regions in CoPu diagrams are restricted by the above relation. The values are further constrained
by (\ref{4vec_pc}) and lie inside hyperbolas within the above light cone boundaries.

Secondly, the CoPu diagram can help us in distinguishing different qubit channels, eg., the unital and nonunital, the incoherent channels, degradable and antidegradable, Pauli channels, etc. These studies unveil very interesting properties of these channels. For example, we find that the qubit incoherent channels can either be unital or nonunital 
with $\vec{\tau} =\{0,0,\tau_z\}$. We also find that all coherence breaking channels has zero coherence. However, this is not usually true for entanglement breaking channels. We observe that the coherence preserving qubit channels have unit coherence. 

Although we mainly focus our study for single qubits, it will be interesting to extend our results for higher dimensions. There are many indications in our work that says one might be able to do it. We hope our findings 
will open up a new direction of studies towards the resource theory of coherence.

\emph{Note:} We notice a work \cite{overlapWork} appeared in the arXiv on the same day of submission of our work in the arXiv. Although their analysis 
is very different than ours, the definition of coherence for quantum channels has overlap with our method.

\end{document}